# Compaction Self-Assembly of Ultralow-Binder-Content Thermoplastic Composites Based on Lunar Soil Simulant

Kiwon Oh,[1] Tzehan Chen,[1] Rui Kou,[2] Haozhe Yi,[2] Yu Qiao,[1,2,*]

[1] *Program of Materials Science and Engineering, University of California – San Diego, La Jolla, CA 92093, U.S.A.*

[2] *Department of Structural Engineering, University of California – San Diego, La Jolla, CA 92093-0085, U.S.A.*

**Abstract:** In a recent study, we developed ultralow-binder-content (UBC) structural materials based on lunar soil simulant and thermoset binders. In the current research, we investigated thermoplastic binders. Compared to thermosets, advanced thermoplastics could be more UV resistant, more durable, more robust, and recyclable. Our main technology is the compaction self-assembly (CSA). By using only ~4 wt% polyetherketoneketone (PEKK) binder, the thermoplastic-binder UBC composite was stronger than typical steel-reinforced concrete. The CSA operation was separate from the curing process. This study may provide an important in-situ resource utilization method for large-scale construction on Moon.

# 1. Introduction

In a recent study on lunar construction materials (Chen et al., 2015a, 2015b, 2017, 2018), we developed a processing technology to use a small percentage of thermoset binder to produce structural parts based on lunar soil simulant. The thermoset binder can be epoxy or unsaturated polyester resin (UPR), intended to be prepared on and transported from Earth. Its content is only ~4 wt%, nearly one order of magnitude less than the matrix content in conventional composite materials (Oh et al., 2019). The other ~96% is the lunar soil simulant, referred to as the filler. The so-processed ultralow-binder-content (UBC) composite can be stronger than typical steel-reinforced concrete, without using any steel rebars. The key concept is the compaction self-assembly (CSA). The filler and the binder are first premixed in a regular mechanical mixer. Since the binder is insufficient to wet the all the filler grains, the mixture is quite sandy, and the binder nonuniformly disperses as small droplets. A compaction pressure is then applied onto the premixed material. The pressure is in the range from 30 MPa to 200 MPa, depending on the materials system. This pressure is higher than the typical pressure range of compression molding (Suherman et al., 2013). It not only densifies the filler grains and squeezes the binder droplets, but also renders the binder self-assembled into polymer micro-agglomerations (PMA) at the places where filler grains are in contact, thanks to the high capillary pressure in the narrow confinement. As most the PMA are located at the most important load-carrying microstructural sites, the system redundancy is greatly decreased and the structural integrity of the UBC composite can be order-of-magnitude better than conventional composites with the same binder amounts. The CSA process offers a critical in-situ resource utilization method, potentially enabling large-scale construction on Moon for lunar bases, outposts, structural parts of research facilities (such as space telescopes and sensor supports), landing and launch platforms, permanent waste burial containers, protection walls, pavements, among others.

One issue of thermoset binders is related to their relatively poor ultraviolet (UV) resistance (Armstrong et al., 1995). Although their contents in UBC composites are low and the filler grains should have a protection effect, it is desirable to use more durable polymers in the harsh space environment. Compared to thermosets, thermoplastics provide more choices. Many advanced thermoplastics, such as polyimide (PI), polyamide (PA), and polyetherketoneketone (PEKK) (Fig. 1) tend to have a higher UV resistance than epoxy and UPR (Alvino, 1971). They could have longer shelf lives, more flexible molding conditions, higher toughness, and better waterproofness (Sezer Hicyilmaz et al., 2019). Moreover, thermoplastics are recyclable (Zhu et al., 2016). It is particularly relevant to the lunar exploration missions. As old materials are melted and remolded into new parts, less additional shipment would be required from Earth.

In the current research, we used PI, PA, and PEKK as binders to produce thermoplastic-based UBC composites, with JSC-1a lunar soil simulant as the filler. The PA was further enhanced by nanoclays. It has been reported that nanoclay could significantly improve the strength and lower the permeability of PA (García et al., 2007; Isitman et al., 2010; Zhang and Loo, 2008), while the workability might be reduced (Chiou et al., 2006; Liu et al., 2004).

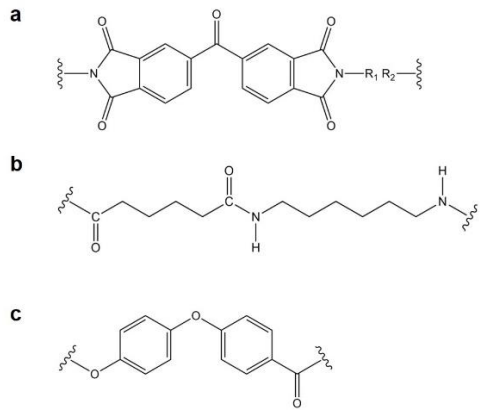

**Fig. 1** Molecular structures of a) PI, b) PA, and c) PEKK.

## 2. Experimental

JSC-1a lunar soil simulant was provided by Orbitec. Polyetherketoneketone (PEKK) was provided by Arkema (Product No. Kepstan 6002PL) in powder form, with the particle size of ~ 50 µm; polyimide (PI) was provided by Evonik Industries (Product No. P84) in powder form, with the particle size ranging from 1-10 µm; polyamide-nanoclay composite (PAnc) was provided by UBE Industries in pellet form, with the grain size around a few mm. The as-received PAnc was ground by an IKA A11 mill for 30 min at 1000 rpm into fine powders 50~200 µm large.

The thermoplastic powders were mixed with about 5 g of air-dried JSC-1a simulant grains in a 50 ml beaker with a lab spatula. The polymer content ranged from 3 wt% to 6 wt%. The filler was not sieve analyzed and its grain size distribution was random. The mixture was moved into a cylindrical load cell. The load cell was made of stainless steel and was 19.05 mm in inner diameter and 50.8 mm in height. It was equipped with a top piston and a bottom piston; both of the pistons were 19.05 mm in diameter and 25.4 mm in height. The top piston was quasi-statically compressed into the load cell by a type-5582 Instron machine, with the loading rate of 0.3 mm/min. The maximum compression pressure ranged from 10 MPa to 350 MPa. After holding the pressure for 5 minutes, the load was reduced to zero at the same rate, and the load cell was moved into Carbolite CTF-1275700 quartz tube furnace. The material was cured at the set point (300~500 °C) for 1 h in a nitrogen protection environment, followed by furnace cooling. The heating rate was 20 °C/min. No external force was applied on the pistons during curing.

The cured material was cut into beam specimens by an MTI high speed diamond cut-off saw. The surfaces were polished by 320 grit sand papers. The flexure strength was measured in a three-point bending setup on the type-5582 Instron machine, in the displacement control mode.

The testing temperature was -190 °C, 25 °C, or 120 °C. The low temperature was achieved by using a cold bath of a mixture of liquid nitrogen and ethanol. The high temperature was achieved by using a heating band, monitored by a type-K thermocouple. The crosshead speed was set to 3.0 mm/min and the span length, $L$, was 19.05 mm. The flexural strength is defined as $R = \frac{3}{2}\frac{P_{\max}L}{bd^2}$, where $P_{\max}$ is the maximum load, $b$ the sample width, and $d$ is the sample height. For each condition, at least three nominally same samples were tested. The fractured surfaces were observed under a scanning electron microscope (SEM).

## 3. Result and discussions

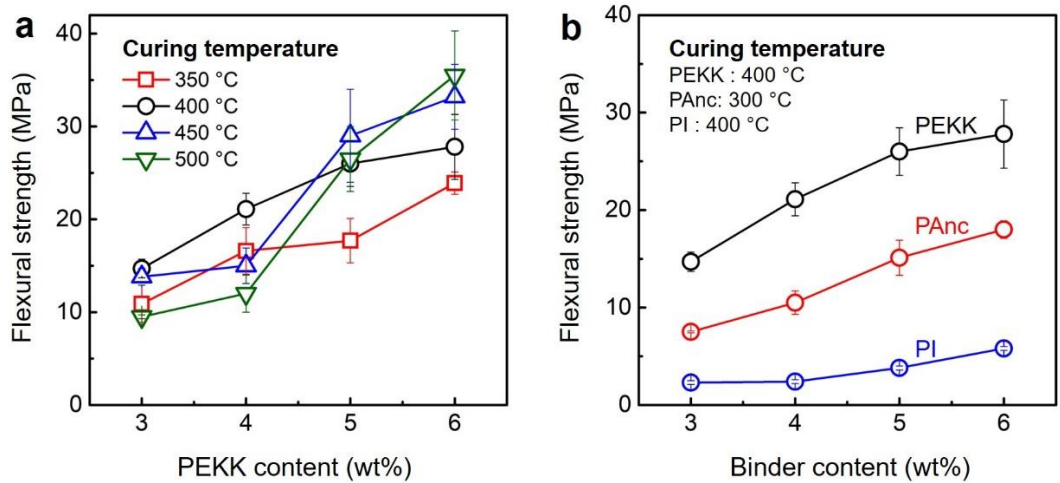

**Fig. 2** The effects of a) the curing temperature and b) the thermoplastic binder content on the flexural strength. The compaction pressure is 350 MPa.

Fig. 2a shows the flexural strengths ($R$) of PEKK-binder UBC composites of various PEKK contents ($c$) and curing temperatures ($T_c$). Clearly, $R$ increases with $c$, as expected for the low binder range under investigation. With only 3 wt% of PEKK, the flexural strength reaches 10~15 MPa, comparable with that of typical steel-reinforced concrete (Wight and Macgregor, 2016). When $c = 6$ wt%, the flexural strength can be as high as ~35 MPa, comparable with the compressive strength of typical concrete (Belaoura, 2010). The curing temperature has a significant effect on the strength. Fig. 3 depicts the CSA mechanism. It is quite different from the CSA of thermoset binders (Chen et al., 2017, 2018). A thermoset binder is in liquid phase when the compaction pressure is applied and therefore, the binder droplets are directly driven by the global densification pressure and the local capillary pressure. In our experiment on the thermoplastic binders, the binder is in solid form during compaction. The binder powders are dispersed through filler grain deformation, sliding, and rotation. Such a short-range movement is irreversible, remaining unchanged after the compaction pressure is removed. The glass transition temperature ($T_g$) of PEKK is ~160 °C and its melting point ($T_m$) is ~300 °C (Marin-Franch et al., 2002); $T_g$ of the PI is higher, around 330 °C (Ree et al., 1992); $T_m$ of PAnc is ~270 °C (Usuki et al., 2005). The curing temperature is kept well above $T_m$. During curing, the binder powders transform into the liquid phase, forming strong bonds with the adjacent filler surfaces. While the pistons are not pressurized externally, they are self-locked and the material inside the load cell is confined. As the binder volume expands, an inner pressure is built up, driving the binder to the narrowest space, i.e., the contact points of filler particles. Hence, somewhat similar to the thermoset binder, polymer micro-agglomerations (PMA) are produced, which efficiently enhances the overall structural properties. The optimum curing temperature is determined by the binder content, the thermal expansion coefficients of the binder and the filler, the filler grain size and surface roughness, etc. For the PEKK materials system, when the binder content is below 4 wt%, the optimum curing temperature is ~400 °C. When $c$ is 5~6 wt%, a higher curing temperature around 450~500 °C is preferred.

In Fig. 2b, the curing temperature has been optimized for each binder type ($c = 4$ wt%). Similar to Fig. 2a, increasing the binder content is beneficial to the strength, as it should be. Generally, PEKK-based UBC composites are stronger than PI-based specimens, which should be attributed to the higher strength and higher tackiness of PEKK (Chang, 1988). However, although PAnc has the highest strength among the three binders (Agag et al., 2001), its UBC composites are weaker than PEKK-based materials. It may be related to the larger PAnc powder size as well as the larger effective viscosity of PAnc melt (Jang and Wilkie, 2005).

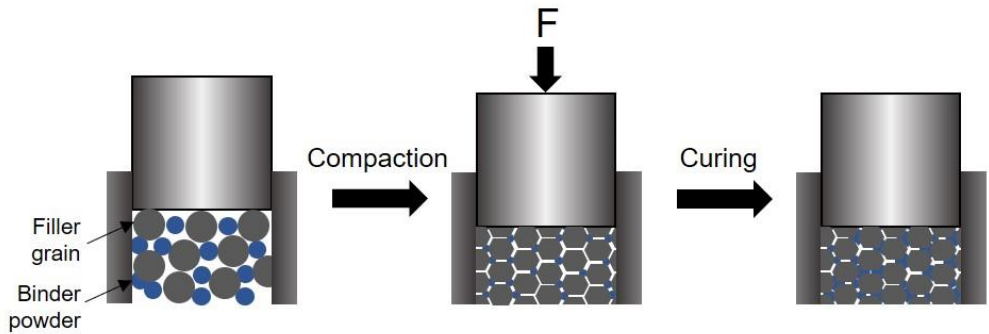

**Fig. 3** Schematic of the CSA process of the thermoplastic-based UBC composite.

Another important factor is the peak compaction pressure, $P_c$. As shown in Fig. 4a, $R$ increases with $P_c$. Without the compaction operation, no structurally integral composite can be fabricated with the ultralow binder amount, and $R$ is essentially zero. When $P_c = 10$ MPa, the flexural strength is ~2 MPa, comparable with regular concrete with a relatively high water/cement ratio (Popovics, 1969). As $P_c$ rises to 100 MPa, $R$ is rapidly improved to ~14 MPa. A larger $P_c$ is still beneficial when it further increases, but the effect becomes less pronounced. As $P_c$ reaches 350 MPa, $R$ is around 20 MPa. Since no liquid form is involved in the compaction process, the influence of $P_c$ should be associated with the plastic deformation of binder powders as well as the internal friction among filler and binder. As shown in the SEM images in Fig. 5, when $P_c$ is relatively low, a relatively large number of filler grains are poorly bonded, resulting in a large defect density and a relatively low strength. As $P_c$ is relatively large, the filler grain deformation and densification is near complete, and the binder dispersion is more homogenous. Because the defect sites are reduced, the strength increases. In general, $R$ depends on the polymer binder content ($c$), the peak compaction pressure ($P_c$), the binder strength ($\sigma_0$), and the flexibility of binder ($1/E$), with $E$ being the modulus of elasticity. Thus, $R = f(P_c, \sigma_0, c, E)$, where $f$ is a certain function. Through dimensional analysis, we have $\frac{R}{c\sigma_0} = f\left(\frac{P_c}{E}\right)$. If $f$ follows a power law,

$$R = \beta_1 c\sigma_0 \left(\frac{P_c}{E}\right)^{\beta_2} \tag{1}$$

where $\beta_1$ and $\beta_2$ are two dimensionless system parameters. For PEKK, $\sigma_0$ is around 130 MPa and $E$ is about 3.5 GPa (ASTM, 2015). When $\beta_1 = 0.12$ and $\beta_2 = 0.46$, Eq.(1) fits with the testing data quite well, as demonstrated by the dashed line in Fig. 4a. Parameter $\beta_2$ is dominated by compaction-related factors, such as the filler and binder grain sizes, the filler-binder internal friction, the deformation and densification of filler, among others. Parameter $\beta_1$ is determined by factors not directly related to compaction, such as the curing temperature and the filler-binder bonding strength. The filler-binder interfacial tension may be assessed as $\gamma = \sqrt{\gamma_f \gamma_b}$ (Wu, 2007), where $\gamma_f$ and $\gamma_b$ are the surface tensions of the filler and the binder, respectively. The effective bonding strength is the difference between $\gamma$ and the summation of $\gamma_f$ and $\gamma_b$. It can be seen that, to achieve a strong bonding, with a given $\gamma_f$ of JSC-1a grains, $\gamma_b$ should be in an intermediate range. If $\gamma_b$ is too large or too small, the wettability of the polymer phase on the simulant grain surfaces would be low.

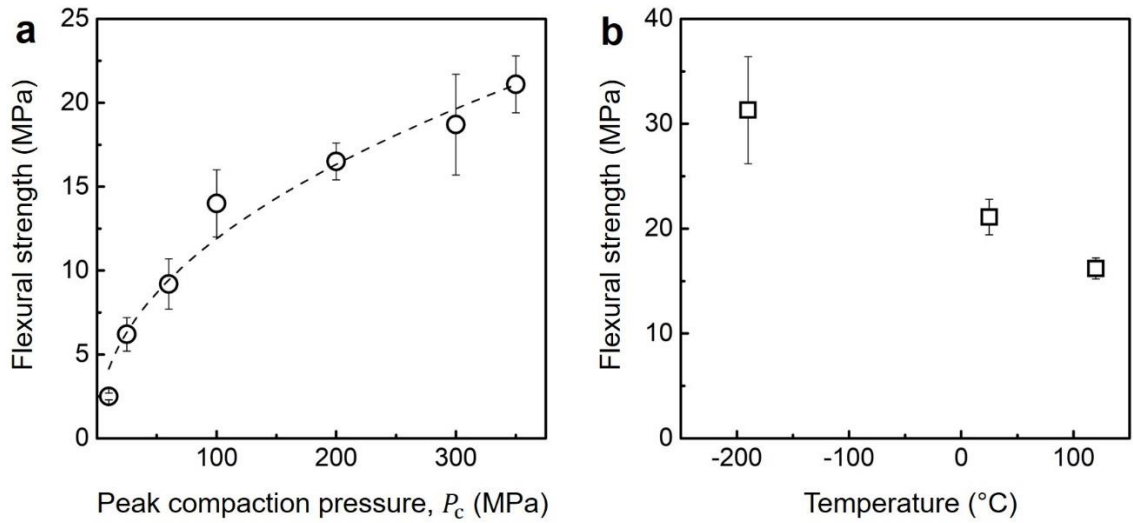

**Fig. 4** The effects of a) the peak compaction pressure and b) the testing temperature on the flexural strength of PEKK-binder UBC composites. The PEKK binder content is 4 wt%; the curing temperature is 400 °C; the compaction pressure is 350 MPa in (b). The dashed line in (a) is calculated from Eq.(1).

Fig. 4b gives the measured strength at various testing temperatures ($T$), in the range relevant to the lunar surface (Sinton, 1962). As $T$ becomes higher, $R$ decreases, following the general trend of thermal effect on polymer strength. At a low temperature around -190 °C, the strength can be as high as 35 MPa, while the data scatter is relatively large, exhibiting typical characteristics of a brittle material. When temperature rises from 20 °C to 120 °C, $R$ decreases from ~20 MPa to ~16 MPa by about 20%, still stronger than many steel-reinforced concretes.

In the majority part of the processing procedure of the thermoplastic-based UBC composites, the binder is in powder form and no heating is involved. Curing is performed in the last step. Hence, the complexity of mixing and handling of polymer melts is avoided. Potential binder loss during compaction is minimized. The burden and energy use of heating and insulation are largely reduced. As heating and compression are de-coupled, the system design and operation is quite straightforward. The thermoplastic-binder composites enjoy many benefits of the thermoset-binder composites. For instance, because the high compaction pressure deforms the filler grains, sieve analysis and control of grain size gradation design is unnecessary.

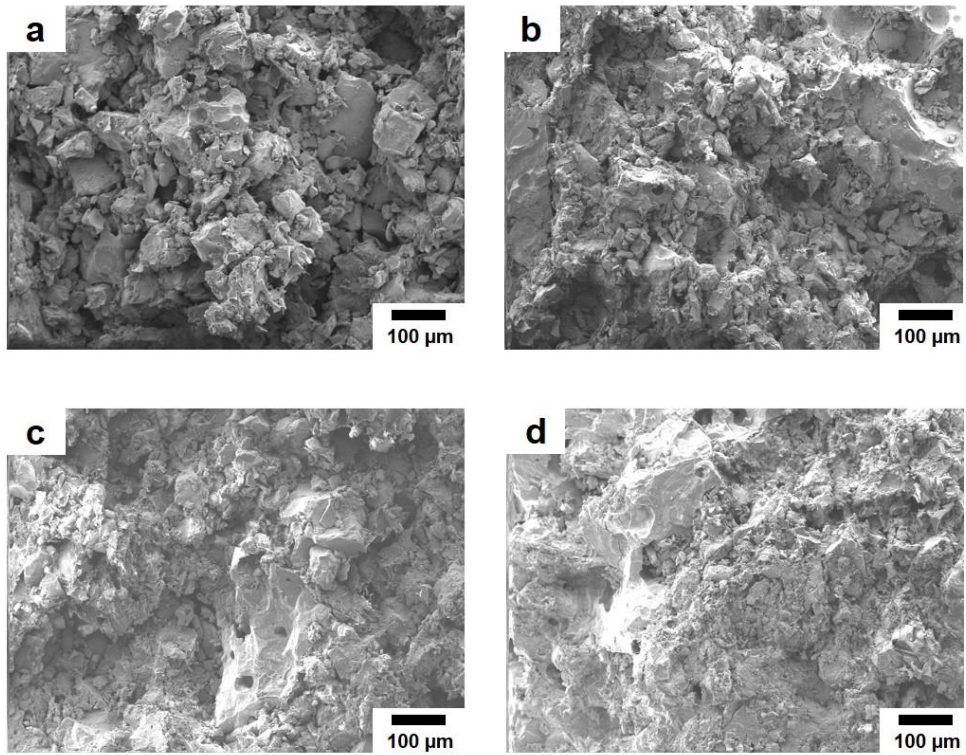

**Fig. 5** Typical SEM images of UBC composites based on 4 wt% PEKK, compacted at a) 60 MPa, b) 100 MPa, c) 200 MPa, and d) 300 MPa. The curing temperature is 400 °C.

## 4. Concluding remarks

In summary, we processed and tested ultralow-binder-content (UBC) composites based on JSC-1a lunar soil simulant and thermoplastic binders. The binders under investigation were polyetherketoneketone (PEKK), polyimide (PI), and polyamide nanoclay composite (PAnc). Through compaction self-assembly (CSA), with only 3~6 wt% binder, the flexural strength of the UBC composites could reach 20~35 MPa. PEKK led to the highest strength. The strength increased with the binder strength and the compaction pressure, and was considerably influenced by the curing temperature. As testing temperature increased, the measured strength was reduced, but still satisfactory. In CSA, the premixed filler grains and binder powders were compacted in solid form. Then, the compacted mixture was cured, during which the binder phase melted and formed bonding with the filler surfaces. This study is important to in-situ resource utilization on Moon, specifically for large-scale construction of lunar bases and research facilities.

**Acknowledgment:** This work was supported by NASA under Grant No. NNX12AI73G and ARPA-E under Grant No. DE-AR0000737 and Grant No. DE-AR0001144.

**References**

Agag, T., Koga, T., Takeichi, T., 2001. Studies on thermal and mechanical properties of polyimide-clay nanocomposites. Polymer (Guildf). 42, 3399–3408. https://doi.org/10.1016/S0032-3861(00)00824-7

Alvino, W.M., 1971. Ultraviolet stability of polyimides and polyamide–imides. J. Appl. Polym. Sci. 15, 2123–2140. https://doi.org/10.1002/app.1971.070150907

Armstrong, R.D., Jenkins, A.T.A., Johnson, B.W., 1995. An investigation into the uv breakdown of thermoset polyester coatings using impedance spectroscopy. Corros. Sci. 37, 1615–1625. https://doi.org/10.1016/0010-938X(95)00063-P

ASTM, 2015. Standard Specification for Polyetherketoneketone ( PEKK ) Polymers for Surgical. Astm 1–6. https://doi.org/10.1520/F2820

Belaoura, M., 2010. Compressive behaviour of concrete at high strain rates. COST ACTION

C26 Urban Habitat Constr. under Catastrophic Events - Proc. Final Conf. 447–451.

Chang, I.Y., 1988. PEKK as a new thermoplastic matrix for high-performance composites. SAMPE Q.;(United States) 19.

Chen, T., Chow, B.J., Qiao, Y., 2015a. Two-step gradation of particle size in an inorganic-organic hybrid. Sci. Eng. Compos. Mater. 22, 643–647. https://doi.org/10.1515/secm-2014-0042

Chen, T., Chow, B.J., Wang, M., Shi, Y., Zhao, C., Qiao, Y., 2015b. Inorganic–Organic Hybrid of Lunar Soil Simulant and Polyethylene. J. Mater. Civ. Eng. 28, 06015013. https://doi.org/10.1061/(asce)mt.1943-5533.0001450

Chen, T., Chow, B.J., Wang, M., Zhong, Y., Qiao, Y., 2017. High-Pressure Densification of Composite Lunar Cement. J. Mater. Civ. Eng. 29, 06017013. https://doi.org/10.1061/(asce)mt.1943-5533.0002047

Chen, T., Chow, B.J., Zhong, Y., Wang, M., Kou, R., Qiao, Y., 2018. Formation of polymer micro-agglomerations in ultralow-binder-content composite based on lunar soil simulant. Adv. Sp. Res. 61, 830–836. https://doi.org/10.1016/j.asr.2017.10.050

Chiou, B. Sen, Yee, E., Wood, D., Shey, J., Glenn, G., Orts, W., 2006. Effects of processing conditions on nanoclay dispersion in starch-clay nanocomposites. Cereal Chem. 83, 300–305. https://doi.org/10.1094/CC-83-0300

García, A., Eceolaza, S., Iriarte, M., Uriarte, C., Etxeberria, A., 2007. Barrier character improvement of an amorphous polyamide (Trogamid) by the addition of a nanoclay. J. Memb. Sci. 301, 190–199. https://doi.org/10.1016/j.memsci.2007.06.018

Isitman, N.A., Aykol, M., Kaynak, C., 2010. Nanoclay assisted strengthening of the fiber/matrix interface in functionally filled polyamide 6 composites. Compos. Struct. 92, 2181–2186.

https://doi.org/10.1016/j.compstruct.2009.09.007

Jang, B.N., Wilkie, C.A., 2005. The effect of clay on the thermal degradation of polyamide 6 in polyamide 6/clay nanocomposites. Polymer (Guildf). 46, 3264–3274. https://doi.org/10.1016/j.polymer.2005.02.078

Liu, T., Tjiu, W.C., He, C., Na, S.S., Chung, T.S., 2004. A processing-induced clay dispersion and its effect on the structure and properties of polyamide 6. Polym. Int. 53, 392–399. https://doi.org/10.1002/pi.1359

Marin-Franch, P., Martin, T., Tunnicliffe, D.L., Das-Gupta, D.K., 2002. PTCa/PEKK piezo-composites for acoustic emission detection. Sensors Actuators, A Phys. 99, 236–243. https://doi.org/10.1016/S0924-4247(01)00789-0

Oh, K., Chen, T., Gasser, A., Kou, R., Qiao, Y., 2019. Compaction self-assembly of ultralow-binder-content particulate composites. Compos. Part B Eng. 175, 107144. https://doi.org/10.1016/j.compositesb.2019.107144

Popovics, S., 1969. Effect of Porosity on the Strength of Concrete. J. Mater.

Ree, M., Chen, K.J., Kirby, D.P., Katzenellenbogen, N., Grischkowsky, D., 1992. Anisotropic properties of high-temperature polyimide thin films: Dielectric and thermal-expansion behaviors. J. Appl. Phys. 72, 2014–2021. https://doi.org/10.1063/1.351629

Sezer Hicyilmaz, A., Altin, Y., Bedeloglu, A., 2019. Polyimide-coated fabrics with multifunctional properties: Flame retardant, UV protective, and water proof. J. Appl. Polym. Sci. 136, 1–10. https://doi.org/10.1002/app.47616

Sinton, W.M., 1962. Temperatures on the lunar surface, in: Physics and Astronomy of the Moon. Academic Press New York, pp. 407–428.

Suherman, H., Sulong, A.B., Sahari, J., 2013. Effect of the compression molding parameters on the in-plane and through-plane conductivity of carbon nanotubes/graphite/epoxy nanocomposites as bipolar plate material for a polymer electrolyte membrane fuel cell. Ceram. Int. 39, 1277–1284. https://doi.org/10.1016/j.ceramint.2012.07.059

Usuki, A., Hasegawa, N., Kato, M., Kobayashi, S., 2005. Polymer-clay nanocomposites, in: Inorganic Polymeric Nanocomposites and Membranes. Springer, pp. 135–195.

Wight, J.K., Macgregor, J.G., 2016. Reinforced Concrete Mechanics and Design. Hoboken.

Wu, S., 2007. Calculation of interfacial tension in polymer systems. J. Polym. Sci. Part C Polym. Symp. 34, 19–30. https://doi.org/10.1002/polc.5070340105

Zhang, X., Loo, L.S., 2008. Morphology and mechanical properties of a novel amorphous polyamide/nanoclay nanocomposite. J. Polym. Sci. Part B Polym. Phys. 46, 2605–2617.

Zhu, Y., Romain, C., Williams, C.K., 2016. Sustainable polymers from renewable resources. Nature 540, 354–362. https://doi.org/10.1038/nature21001